\documentclass[12pt]{amsart}
\usepackage{amsmath,amsthm,amsfonts,amssymb,eucal, hyperref}

\newcommand{\Gad}{\Gamma^{\dagger}}

\newcommand{\oq}{\ {\raise 7pt\hbox{${\scriptstyle\circ}$}}
\kern -7pt{
\hbox{$Q$}}}

\newcommand{\R}{ \mathbb R}
\newcommand{\Q}{ \mathbb Q}
\newcommand{\C}{ \mathbb C}
\newcommand{\N}{ \mathbb N}

\newcommand{\To}{\mathbb T}
\newcommand{\Td}{\mathbb T^{\dagger}}

\newcommand {\GH}{\mathfrak H}

\newcommand {\bx}{\mathbf x}

\newcommand {\be}{\mathbf e}

\newcommand {\bk}{\mathbf k}
\newcommand {\bp}{\mathbf p}

\newcommand {\by}{\mathbf y}

\newcommand {\bn}{\mathbf n}

\newcommand {\bnu}{\boldsymbol\nu}

\newcommand {\bmu}{\boldsymbol\mu}
\newcommand {\bka}{\boldsymbol\kappa}
\newcommand {\bth}{\boldsymbol\theta}

\newcommand {\bxi}{\boldsymbol\xi}

\newcommand{\lu}{\langle}
\newcommand{\ru}{\rangle}

\newcommand{\CO}{\mathcal O}

\newcommand{\1}
{{\,\vrule depth3pt height9pt}{\vrule depth3pt height9pt}
{\vrule depth3pt height9pt}{\vrule depth3pt height9pt}\,}

\newtheorem{thm}{Theorem}[section]
\newtheorem{cor}[thm]{Corollary}
\newtheorem{cla}[thm]{Claim}
\newtheorem{lem}[thm]{Lemma}

\theoremstyle{definition}

\newtheorem{rem}[thm]{Remark}

\numberwithin{equation}{section}

\newcommand{\bee}{\begin{equation}}
\newcommand{\ene}{\end{equation}}
\newcommand{\bees}{\begin{equation*}}
\newcommand{\enes}{\end{equation*}}
\newcommand{\bes}{\begin{split}}
\newcommand{\ens}{\end{split}}

\newcommand{\bet}{\begin{thm}}
\newcommand{\ent}{\end{thm}}
\newcommand{\bel}{\begin{lem}}
\newcommand{\enl}{\end{lem}}
\newcommand{\bec}{\begin{cor}}
\newcommand{\enc}{\end{cor}}
\newcommand{\becl}{\begin{cla}}
\newcommand{\encl}{\end{cla}}
\newcommand{\bep}{\begin{proof}}
\newcommand{\enp}{\end{proof}}
\newcommand{\ber}{\begin{rem}}
\newcommand{\enr}{\end{rem}}
\newcommand{\ep}{\varepsilon}
\newcommand{\la}{\lambda}

\newcommand{\de}{\delta}
\newcommand{\De}{\Delta}
\newcommand{\al}{\alpha}
\newcommand{\Z}{\mathbb Z}
\newcommand{\Ga}{\Gamma}

\makeatletter
\def\square{\RIfM@\bgroup\else$\bgroup\aftergroup$\fi
  \vcenter{\hrule\hbox{\vrule\@height.6em\kern.6em\vrule}\hrule}\egroup}
\makeatother

\begin{document}

\hoffset -4pc

\title[Perturbation theory]
{Perturbation theory for spectral gap edges of 2D periodic Schr\"odinger operators}
\author[L. Parnovski \& R. Shterenberg]
{Leonid Parnovski \& Roman Shterenberg}
\address{Department of Mathematics\\ University College London\\
Gower Street\\ London\\ WC1E 6BT\\ UK}
\email{Leonid@math.ucl.ac.uk}
\address{Department of Mathematics\\ University of Alabama at Birmingham\\ 1300 University Blvd.\\
Birmingham AL 35294\\ USA}
\email{shterenb@math.uab.edu}


\date{\today}

\begin{abstract}
We consider a two-dimensional periodic Schr\"odinger operator $H=-\Delta+W$ with $\Gamma$ being the lattice of periods. We investigate the structure of the edges of open gaps in the spectrum of $H$. We show that under arbitrary small perturbation $V$ periodic with respect to $N\Gamma$ where $N=N(W)$ is some integer, all edges of the gaps in the spectrum of $H+V$ which are perturbation of the gaps of $H$ become non-degenerate, i.e. are attained at finitely many points by one band function only and have non-degenerate quadratic minimum/maximum. We also  discuss this problem in the discrete setting and show that changing the lattice of periods may indeed be unavoidable to achieve the non-degeneracy. 
\end{abstract}



\maketitle
\vskip 0.5cm

\renewcommand{\uparrow}{{\mathcal {L F}}}
\newcommand{\ssharp}{{\mathcal {L E}}}
\renewcommand{\natural}{{\mathcal {NR}}}
\renewcommand{\flat}{{\mathcal R}}
\renewcommand{\downarrow}{{\mathcal {S E}}}

\section{Introduction}

Let 
\bee
H=-\De+W
\ene
be a Schr\"odinger operator in $\R^d$, $d\ge 2$, with a smooth periodic potential $W=W(\bx)$. Let $\Ga$ be its lattice of periods and $\Gad$ be the dual lattice. 
We put $\To:=\R^d/\Ga$ and $\Td:=\R^d/\Gad$. 
It is known \cite{ReSi} that the spectrum of $H$ is 
\bee
\sigma(H)=[\la_0,+\infty)\setminus (\sqcup_{m=1}^n (\mu_{m,-},\mu_{m,+})),
\ene 
where the non-intersecting intervals $(\mu_{m,-},\mu_{m,+})$ are called 
{\it the gaps}. There are finitely many of them, and for small $V$ there are no gaps at all, \cite{Sk,Pa}. We are interested in the behaviour of the spectrum of $H$ near the spectral edges $\mu_{m,\pm}=\mu_{m,\pm}(H)$. More precisely, consider the Floquet-Bloch decomposition of $H$ into the direct integral:
\bee\label{direct}
H=\int^{\oplus}_{\Td}H(\bk)d\bk
\ene 
(see Section \ref{facts} for more details) and denote by $\{\la_j(\bk)\}_{j=0}^{\infty}$ the collection of eigenvalues of $H(\bk)$ (in non-decreasing order, taking multiplicities into account). 
Then each function $\la_j$ (called the {\it Bloch function}) is smooth, at least outside the values of $\bk$ where the values of two such functions coincide. Then 
\bee
\sigma(H)=\cup_{\bk}\cup_j\la_j(\bk).
\ene
Therefore, for any spectral edge $\mu_+=\mu_{m,+}$ for some $m$ there is 
a point $\bk_0$ and an index $j=j(\mu_+)$ such that $\la_j(\bk_0)=\mu_+$, and $\bk_0$ is a point of local (even global) minimum of $\la_j$ (similarly, for any edge $\mu_-$ there is a Bloch function $\la_j$, $j=j(\mu_-)$, for which $\mu_-$ is a maximal value). 
We are interested in the behaviour of the function $\la_j$ in a neighbourhood of $\bk_0$. One may expect that for each spectral edge $\mu_\pm$ the following properties hold, at least generically: 

A. No other Bloch function takes value $\mu_\pm$, meaning that if for some $l$ and $\bk\in\Td$ we have $\la_l(\bk)=\mu_\pm$, then $l=j$. 

B. The set of points 
\bee
S=S(\mu_\pm):=\{\bk_0\in\Td,\ \la_j(\bk_0)=\mu_\pm\}
\ene
is finite.

C. The quadratic form of $\la_j$ around each critical point $\bk_0\in S$ is non-degenerate, meaning that 
\bee\label{quadratic0}
\la_j(\bk)=\mu_{\pm} \pm [A(\bk-\bk_0)] (\bk-\bk_0)+o(|\bk-\bk_0|^2)
\ene
with positive definite matrix $A$. 

If these properties hold, we say that the spectral edge $\mu_\pm$ is non-degenerate. In the physical papers it is often assumed that generic Schr\"odinger operators have non-degenerate spectral edges. For example, 
in solid state physics, the
tensor of effective masses is essentially defined as the inverse of the matrix $A$ from 
\eqref{quadratic0} (see e.g. \cite{A}). This  definition 
makes  sense  only  if  all three conditions are satisfied. In one-dimensional situation the spectral edges are always non-degenerate, see e.g. \cite{ReSi}. The bottom of the spectrum is known to be non-degenerate in all dimensions, see \cite{KiSi} (but the same cannot be said about a magnetic Schr\"odinger operator, see \cite{Sh}, where an example of a magnetic operator the bottom of whose spectrum does not satisfy Condition C is given).   It  is  commonly  believed  that in multidimensional case ($d\ge 2$) the  spectral
gap edges are non-degenerate for generic potentials, see, for example, \cite{KuPi} and \cite{Ku}, where additional references are given. 

Property A has been established to hold generically in \cite{KlRa}. In a recent paper \cite{FiKa} Property B has been proved for {\bf {all}} (not just generic) operators if $d=2$. It is not known whether Property B holds even generically in higher dimensions. The remarkable simple example discovered recently by N.Filonov \cite{FiKa} shows that for discrete periodic Schr\"odinger operators Property B does not hold, not even generically: there is a discrete periodic Schr\"odinger operator $\hat H$ for which the set $S$ corresponding to a spectral edge consists of two intervals, and the same holds for all operators close to $\hat H$. It turns out, however, that this feature of $\hat H$ is generically destroyed if we perturb $\hat H$ by a potential with a smaller lattice $\tilde \Ga\subset\Ga$, where $\Ga$ is the initial lattice of periods of $\hat H$. 
We discuss this and related results in Section \ref{example}. 

The main result of our paper concerns Property C in the two-dimensional case. We will prove that all spectral edges of $H$ can be made non-degenerate by perturbing it with arbitrarily small periodic potential $V$,  with a smaller lattice of periods $\tilde \Ga\subset\Ga$. Namely, we will  prove the following result:

\bet\label{main}
Let $W=W(\bx)$, $\bx\in\R^2$, be a smooth function periodic with respect to some lattice $\Gamma$. Then for every $\epsilon>0$ there exist $N=N(W,\epsilon)\in\N$ and a potential $V(W,\epsilon)$ periodic with respect to $\tilde\Ga:=N\Gamma$ and satisfying $\|V\|_\infty<\epsilon$ 
such that the 
following property holds. Suppose, $(\mu_{m,-}(H_1),\mu_{m,+}(H_1))$
is a spectral gap of the 
operator
$H_1:=-\De+W+V$ with $\mu_{m,+}(H_1)-\mu_{m,-}(H_1)>\epsilon$. Then the edges $\mu_{m,\pm}$ are non-degenerate. 
\ent

\ber 

1. 
As previous paragraph (and Section \ref{example}) show, decreasing the lattice of periods to achieve non-degeneracy may be necessary in the discrete case; we do not know whether it is possible to make spectral edges non-degenerate by perturbing $H$ with a small potential with the same lattice of periods $\Ga$, nor do we know whether this result holds in higher dimension (a substantial part of our proof is based on the fact that Property B holds for all, not just generic operators, and there are no high-dimensional analogues of \cite{FiKa} known so far). 
 
2. There are two problems we have to deal with when increasing the lattice of periods. The first one is that the perturbed operator $H_1$ can have more spectral  gaps than $H$. If the lattice of periods of $V$ is $\Ga$, the number of new gaps has an upper bound depending only on $H$ and $||V||_{\infty}$. 
If the lattice of periods of $V$ is $N\Ga$, then the number of new gaps may also depend on $N$, and our construction of the perturbation $W$ 
has no control on the size of $N$. Therefore, what can happen in principle is the following. We introduce a perturbation $V_1$ periodic with respect to $N_1\Ga$ so that $H+V_1$ has edges of `old' gaps non-degenerate, but some `new' gaps (of very small length) may appear. Then we may deal with these gaps by adding another, even smaller perturbation $V_2$ with lattice of periods $N_2\Ga$, but a further set of `new new' gaps may be opened, etc. We do not have control over whether this process can last indefinitely long. Therefore, in our theorem we can guarantee only that all the edges of `old' gaps (i.e. gaps the length of which is not small) become non-degenerate. 

3. Another problem of dealing with the perturbations with increasing lattice of periods is the stability issue. It follows from the standard perturbation theory that the non-degeneracy of the spectral edges is stable under further perturbations with the same lattice of periods. More precisely, let $\epsilon,\ N,\ V$ be as in Theorem~\ref{main} then the conclusion of the theorem holds for operator $H_2=-\De+W+V+Q$ with any smooth potential $Q$ periodic with respect to $N\Gamma$, provided $\|Q\|_\infty\leq\delta$ with $\delta=\delta(W,V,N,\epsilon)$ being sufficiently small. We, however, cannot guarantee the same result if the lattice of periods of $Q$ can increase further and become $NM\Ga$ with $M\in\N$ (we can probably achieve this stability only at the edges of the `old' gaps by introducing an extremely weird-looking norm in the class of all periodic operators with lattices being a sub-lattice of $\Ga$, but the proof of corresponding statement is rather long and unhelpful, so we do not include it here). 

4. As can be seen from the construction, the potential $V$ is a finite  trigonometric polynomial. As a result, we can choose our perturbation satisfying $\|V\|_s<\epsilon$, where $s$ is a fixed real number and $||\cdot||_s$ is a Sobolev norm.

5. The intuition behind the behaviour of the Bloch functions by the perturbations with increasing lattices of periods comes from studying the almost-periodic Schr\"odinger operators. 

6. The same conclusions will hold if we consider a more general class of unperturbed operators, say the periodic magnetic Schr\"odinger operators (or even periodic second order coefficients); effectively, the only property we need from a class of operators we consider is the finiteness of the set $S$, see \cite{FiKa}.

\enr

The rest of the paper is constructed in the following way. In section \ref{facts}, we introduce the necessary notation and discuss how the decomposition \eqref{direct} changes when we increase the lattice of periods $\Ga$. In Section \ref{3} we prove Theorem \ref{main} and also give a simple proof of Property A (proved originally in \cite{KlRa}). Finally, in Section \ref{example}, we discuss the discrete situation when Property B is violated.  


\subsection*{Acknowledgments}
We are grateful to Nikolay Filonov, Ilya Kachkovskiy and Peter Kuchment for useful discussions and to the referee for several important suggestions. 
We also would like to thank the Isaac Newton Institute for Mathematical Sciences for its hospitality during the programme `Periodic and Ergodic Spectral Problems' supported by EPSRC grant EP/K032208/1.
The visit of LP to the Newton Institute was partially supported by the Simons Foundation. 
The research of LP was partially supported by the EPSRC grant EP/J016829/1; RS was partially supported by the NSF grant CCF-1527822. 

\section{General facts}\label{facts}

Suppose, 

\bee\label{def1}
H=-\De+W
\ene

is a Schr\"odinger operator with periodic potential $W$ acting in $\R^d$. For simplicity we assume $W$ to be smooth, but in fact we do not require this assumption for our results. 
Let $\Ga$ be its lattice of periods and $\Gad$ be the dual lattice. 
We put $\To:=\R^d/\Ga$ and $\Td:=\R^d/\Gad$. 
For each quasi-momentum $\bk\in\Td$ we denote by $H(\bk)$ the fibre operator of $H$ corresponding to $\bk$ so that 
\bee
H=\int^{\oplus}_{\Td}H(\bk)d\bk.
\ene

The domain of $H(\bk)$ consists of functions from $H^2(\To)$ satisfying $\bk$-quasi-periodic boundary conditions; let us denote this space by $H^2(\To;\bk)$. The action of $H(\bk)$ (considered as an unbounded operator acting in $L_2(\To)$) is given by the formula 

\bee\label{def2}
H(\bk)f=-\De f+Wf, \ \ f\in H^2(\To;\bk).
\ene

The other, more convenient way of defining these operators is the following.
First, we put
\bee
\be_{\bxi}(\bx):=e^{i\bxi\bx}, \ \ \bxi,\bx\in\R^d.
\ene

We then denote by $H^s(\R^d;\Ga;\bk)$ the space of all infinite Fourier series of the form 

\bee\label{f}
f=\sum_{\bth\in\Gad}a_{\bth}\be_{\bth+\bk},
\ene

where $a_{\bth}=\lu f,\be_{\bth+\bk}\ru_{L^2(\To)}|\To|^{-1}\in\C$ satisfy 

\bee\label{norm}
\sum_{\bth\in\Gad}|a_{\bth}|^2(|\bth+\bk|^2+1)^s<+\infty.
\ene

Finally, we say that $H^s(\To;\bk)$ is the restriction of $H^s(\R^d;\Ga;\bk)$ to the torus $\To$,  
and the LHS of \eqref{norm} multiplied by $|\To|$ defines the square of the norm of $f$ in  $H^s(\To;\bk)$. It will be convenient to define $L^2(\R^d;\Ga;\bk)$
as the collection of functions of the form \eqref{f} with 

\bee\label{norm1}
\sum_{\bth\in\Gad}|a_{\bth}|^2<+\infty 
\ene
and $L^2(\To;\bk)$ as the restriction of $L^2(\R^d;\Ga;\bk)$ to the torus $\To$. Note that when we change $\bk$, the space $L^2(\R^d;\Ga;\bk)$ does not change as the collection of elements (but the form in which we write these elements does change), whereas $H^2(\R^d;\Ga;\bk)$ changes with $\bk$. 

Given another function 

\bee\label{g}
g=\sum_{\bth\in\Gad}g_{\bth}\be_{\bth+\bk}
\ene
from $L^2(\To;\bk)$, we obviously have 

\bee
\lu f,g\ru_{L^2(\To)}=|\To|\sum_{\bth\in\Gad}a_{\bth}\overline {b_{\bth}}.
\ene

Suppose, the Fourier decomposition of $W$ has the following form:

\bee\label{W}
W=\sum_{\bth\in\Gad}w_{\bth}\be_{\bth}.
\ene

Then the action of $H(\bk)$ on the function $f$ is given by:

\bee
H(\bk)f=\sum_{\bth\in\Gad}\left[a_{\bth}|\bth+\bk|^2+\sum_{\bth_1\in\Gad}a_{\bth_1}w_{\bth-\bth_1}\right]\be_{\bth+\bk}.
\ene

We denote by $\{\la_j(\bk)\}$ ($j=0,1,\dots$) the collection of eigenvalues of $H(\bk)$ (counted with multiplicities; it will be convenient from now on to stop assuming that $\la_j(\bk)$ are listed in the increasing order) 
and by $\psi_j=\psi_j(\bk)=\psi_j(\bk;\bx)$ corresponding orthonormal eigenfunctions. We will also assume (as we can without loss of generality) that $\la_j$ are piecewise continuous. We also denote
\bee
\lu \psi_j(\bk),\be_{\bth+\bk}\ru_{L^2(\To)}/|\To|=:\hat\psi_j(\bth;\bk),
\ene
so that 
\bee\label{psi}
\psi_j(\bk)=\sum_{\bth\in\Gad}\hat\psi_j(\bth;\bk)\be_{\bth+\bk}
\ene
and
\bee\label{e}
\be_{\bth+\bk}=|\To|\sum_{j}\overline{\hat\psi_j(\bth;\bk)}\psi_j(\bk).
\ene
These formulas also show that 
\bee\label{psi1}
|\To|\sum_{\bth\in\Gad}\hat\psi_j(\bth;\bk)\overline{\hat\psi_m(\bth;\bk)}=\de_{jm}
\ene
and 
\bee\label{psi2}
|\To|\sum_{j}\hat\psi_j(\bth_1;\bk)\overline{\hat\psi_j(\bth_2;\bk)}=\de_{\bth_1\bth_2}. 
\ene

Now we discuss how this decomposition changes when we increase the lattice (in the sense that we increase the size of a cell of the lattice). Let $N$ be a natural number and put $\tilde\Ga:=N\Ga$. Then $(\tilde\Ga)^{\dagger}=\Gad/N$. We also put $\tilde \To:=\R^d/\tilde\Ga$ and $\tilde \Td:=\R^d/\tilde\Gad$. The quotient group $(\tilde\Ga)^{\dagger}/\Gad$ consists of $M:=N^d$ elements; let us denote by $\{\bp_1=0,\dots,\bp_M\}$ representatives of the elements of this group in $\tilde\Gad$. Then each element of $\tilde\Gad$ has a unique representation 
in the form $\bp_l+\bth$, $\bth\in\Gad$. Moreover, every element 
$\bk\in\Td$ can be uniquely written as $\bp_l+\bka$, $l=1,\dots M$ and $\bka\in\tilde\Td$. In this case we say that $\bka=\kappa(\bk)$ and $l=L(\bk)$. 
This defines a mapping $L:\Td\to \{1,\dots,M\}$ and a mapping $\kappa:\Td\to\tilde\Td$; each point $\bka\in\tilde\Td$ will have exactly $M$ pre-images under mapping $\kappa$. 
Sometimes we will call coordinate $\bk$ {\it the old quasimomentum} and $\bka$ {\it the new quasimomentum}.

Suppose, $\bka\in\tilde\Td$. Then the space $L^2(\tilde\To;\bka)$ consists of all the expansions of the form 
\bee\label{ft}
f=\sum_{\tilde\bth\in\tilde\Gad}a_{\tilde\bth}\be_{\tilde\bth+\bka}=
\sum_{l=1}^M \sum_{\bth\in\Gad}a_{\bp_l+\bth}\be_{\bp_l+\bth+\bka}. 
\ene
Obviously, we can treat the RHS of \eqref{ft} as a sum of functions from 
$L^2(\R^d;\Ga;\bk_l)$,  
where $\bk_l$ runs over all the pre-images of $\bka$ under the mapping $\kappa$. 

Suppose, $f\in L^2(\To;\bk)$. Then expansion \eqref{f} can be looked upon as the element from the space $L^2(\tilde\To;\bka)$ with $\bka=\kappa(\bk)$. 
This defines a mapping $F: L^2(\To;\bk)\to L^2(\tilde\To;\kappa(\bk))$; obviously, this mapping maps also $H^s(\To;\bk)$ to $H^s(\tilde\To;\kappa(\bk))$.

\bel
Suppose, $\bka$ is fixed and $\bk_1$ and $\bk_2$ are two different 
pre-images of $\bka$ under the mapping $\kappa$. Suppose, $f_j\in L^2(\To;\bk_j)$, $j=1,2$. Then $F(f_1)$ is orthogonal to $F(f_2)$ (obviously, we talk about $L^2(\tilde\To)$-inner product here).    
\enl
\bep
The proof is straightforward. We have: 
\bee\label{fj}
f_j=\sum_{\bth\in\Gad}a_{\bth}^j\be_{\bth+\bk_j},
\ene
so 
\bee
\bes
\lu F(f_1),F(f_2)\ru_{L^2(\tilde\To)}&=\int_{\tilde\To}
\sum_{\bth_1\in\Gad}a_{\bth_1}^1\be_{\bth_1+\bk_1}(\bx)
\sum_{\bth_2\in\Gad}\overline{a_{\bth_2}^2}\be_{-\bth_2-\bk_2}(\bx)d\bx\\
&=
\sum_{\bth_1\in\Gad}\sum_{\bth_2\in\Gad}
a_{\bth_1}^1\overline{a_{\bth_2}^2}\int_{\tilde\To}
\be_{\bth_1-\bth_2+(\bk_1-\bk_2)}(\bx)d\bx=0,
\end{split}
\ene
since $\bth_1-\bth_2+(\bk_1-\bk_2)\ne 0(\mathrm{mod}\ \tilde\Gad)$. 
\enp
\bec\label{cor:1}
Suppose, $f\in L^2(\To;\bk)$ and $l=2,\dots,M$. Then $\be_{\bp_l}F(f)$ is orthogonal to $F(f)$ in $L^2(\tilde\To)$. 
\enc

These considerations have the following implications to the spectral decomposition of operator $H$ considered as periodic operator with lattice of periods $\tilde\Ga$. Suppose, $\bka\in\tilde\Td$. Then functions 
$\{F(\psi_j(\bka+\bp_l;\cdot))/\sqrt{M}\}$, $j=0,1,\dots$; $l=1,\dots,M$ form an orthonormal basis in $L^2(\tilde\To;\bka)$. The matrix of $H(\bka)$ (considered as an operator acting in $L^2(\tilde\To;\kappa(\bk))\ $) in this basis is diagonal with $\{\la_j(\bka+\bp_l)\}$ standing on the diagonal. We denote 
\bee
\phi_{j,l}(\bka):=F(\psi_j(\bka+\bp_l;\cdot))/\sqrt{M}.
\ene

\section{Description of the approach: main tools}\label{3}
Suppose that $d=2$ and our operator $H$ has a gap $(\mu_-,\mu_+)$ in its spectrum. 

{\bf Definition.} We say that $\mu_+$ (resp. $\mu_-$) is a non-degenerate end of the spectral gap, if  there are finitely many points 
$\bk_0,\bk_1,\dots,\bk_n\in\To$ such that $\la_j(\bk_l)=\mu_+$ (resp. $\la_j(\bk_l)=\mu_-$) for some $j$ (not depending on $l$), for any $m\ne j$ the equation 
$\la_m(\bk)=\mu_+$ (resp. $\la_m(\bk)=\mu_-$) has no solutions $\bk\in\To$, 
 and in the neighbourhood of each $\bk_l$ the function $\la_j$ behaves quadratically:
\bee\label{quadratic}
\la_j(\bk)=\mu_{\pm} \pm [A(\bk-\bk_l)] (\bk-\bk_l)+o(|\bk-\bk_l|^2)
\ene
as $\bk\to\bk_l$ 
for some positive definite matrix $A=A_l$.

We want to prove that, generically, each end of the gap is non-degenerate. For the sake of definiteness, we will be working with the top end of the gap, but the proof will easily extend to the bottom end of the gap and similar results will hold for $\mu_-$. 
Denote 
\bee
S=S(H):=\{\bk:\exists j, \la_j(\bk)=\mu_+\}.
\ene
A recent result of Filonov and Kachkovskiy \cite{FiKa} shows that 
the set $S$ is finite. 

Suppose, $\bnu$ is arbitrary non-zero vector from $\R^d$.
We denote $\de:=|\bnu|$, $\bn:=\bnu\de^{-1}$, and $v:=\be_{\bnu}+\be_{-\bnu}$. 
The perturbations we consider will be of type   
$
H_{\ep}:=H+\ep V,
$
where $V$ is the operator of multiplication by $v$ and $\ep>0$ a small parameter. 
We will always assume that that $\de$ is smaller than the distances between different points in $S$ and that the perturbed operator $H_{\ep}$ is still periodic with, possibly, a
new lattice of periods $\tilde\Ga\subset\Ga$, i.e. that $\bnu$ is a rational multiple of a vector from $\Gad$: $\bnu\in\Q\Gad$. After adding $\ep V$ to our initial operator $H$, the new operator becomes periodic with respect to a new lattice of periods $\tilde\Ga$; we define this lattice as the lattice dual to $\tilde\Gad$ -- the lattice generated by $\Gad$ and $\bnu$. Note that the new lattice $\tilde\Gad$ contains more elements than the old one and therefore it may happen that some points from $S$ (different modulo old lattice $\Gad$) become `glued together' after introducing the shift by $\bnu$, i.e. there may be two points $\bk_1,\bk_2\in S$ so that $\bk_1+n\bnu=\bk_2$ $(\mod\Gad)$ for some integer $n$. We impose an additional condition  that such `gluing' does not occur. Later, we will explain why such a choice of $\bnu$ is always possible (see Subsection \ref{choice}).

For sufficiently small $\ep$ the operator $H_{\ep}$ will have a gap $(\mu_{\ep -},\mu_{\ep +})$, where $|\mu_{\ep\pm}-\mu_{\pm}|=o(1)$ as $\ep\to 0$. The set $S_{\ep}:=S(H_{\ep})$ of quasimomenta where one of eigenvalues of $H_{\ep}$ equals 
$\mu_{\ep +}$ lies in a small neighbourhood of $S$. From now on we will always assume for simplicity 
that $\mu_+=0$, but we will nevertheless often write $\mu_+$ to emphasise that we are at the upper edge of the spectral gap. 

\ber
In this section, as well as in the rest of the paper, we will work only with the dual space -- the space where the quasimomentum $\bk$ is located. We, thus, no longer need letters $\bx$ or $\by$ to denote the original spacial variables. Therefore, we sometimes will be using this fact and introduce new coordinates denoted by $x$ or $y$, etc. in the {\bf {dual}}
space (the space where the quasimomenta live). 
\enr

We will also need the following simple result which easily follows from the 
analytical perturbation theory, see e.g. \cite{Ka}. 
Suppose, $\la_n(\bk_0)$ is a simple eigenvalue of $H(\bk_0)$. Then for $\bk$ near $\bk_0$  there exists a unique eigenvalue of $H(\bk)$ close to $\la_n(\bk_0)$; denote it 
$\la_n(\bk)$. We also can find a neighbourhood of $\bk_0$ (denoted by  
$\CO(\bk_0)$) such that $\la_n(\bk)$
is a simple eigenvalue of $H(\bk)$ whenever $\bk$
is inside the closure of $\CO(\bk_0)$. Obviously, the same will hold for $H'(\bk):=H(\bk)+V$, where $||V||$ -- the $L^{\infty}$-norm of $V$  is sufficiently small; we denote the corresponding eigenvalue by $\la_n'(\bk)$. In these notations we have:

\bel\label{continuity}
Suppose that 
$x,y$ are some orthogonal coordinates around $\bk_0$, $l,m\ge 0$ are integers  and $\ep>0$ is given. Then there  
exist a real number 
$\eta=\eta(\ep,\CO(\bk_0))>0$ such that if $||V||<\eta$ and $\bk\in\CO(\bk_0)$, then we have 
\bee\label{Kato}
|\frac{\partial^{l+m}(\la_n'-\la_n)}{\partial^l x\partial^m y}(\bk)|<\ep.
\ene
\enl
\bep
Indeed, \eqref{Kato} with $l=m=0$ is an immediate consequence of analytic perturbation theory and holds in a slightly bigger neighbourhood than $\CO(\bk_0)$. This together with analyticity of 
$(\la_n'-\la_n)$ implies \eqref{Kato} for arbitrary $l,m$. 
\enp

This lemma shows that in order to prove our main result, it is enough to prove 
the following statement:
\bet
Suppose, $H$ is a periodic operator with $[\mu_-,\mu_+]$ being its spectral gap. Suppose, $\ep\in(0,(\mu_+-\mu_-)/2)$ and $s$ are two fixed real numbers. Then there exists a periodic potential $V$ with $H^s$-norm smaller than $\ep$ such that $H+V$ is periodic  with a spectral gap $[\mu_-',\mu_+']$ with $|\mu_{\pm}'-\mu_{\pm}|<\ep$ and both spectral ends $\mu_{\pm}'$ being non-degenerate. 

\ent
 
Indeed, if the spectrum of $H$ has several gaps in it, we first make the edges of the first gap non-degenerate with the help of sufficiently small perturbation $V_1$. Then, we deal with the edges of the second gap by introducing a (even smaller) second perturbation $V_2$. Note that Lemma \ref{continuity} implies that if $V_2$ is small enough, then the edges of the first gap remain non-degenerate. Of course, while we are doing this, we may open `new' gaps, but their length will be smaller than the sum of the norms of our perturbations $V_j$. 
 It remains  to notice that the number of gaps of $H$ is finite, see \cite{PS} or \cite{Pa}. 
 
\subsection{Description of the approach: The main idea}

Let us for now assume for simplicity that 
there is a unique point 
$\bk_0$ such that 
\bee\label{bk1}
\la_j(\bk_0)=\mu_+
\ene
for some $j$, but \eqref{quadratic} does not hold. Results of \cite{KlRa} show that \eqref{bk1} generically can hold only for one value of $j$ (in the next subsection we will give a short proof of this); we will assume WLOG that $j=0$, 
so that $\la_0(\bk_0)=\mu_+$ (remember that the labelling of eigenvalues is not necessarily done in increasing order, but   
$\la_0(\bk)$ continuously depends on $\bk$ in some neighbourhood of $\bk_0$). We will never use this specific labelling and have chosen it only for convenience.  
We also assume that there is a vector $\bn$ of unit length such that
$\frac{\partial^2\la_0}{\partial \bn^2}(\bk_0)=0$ (otherwise there is nothing left to do). This means that 
\bee\label{alpha}
\la_0(\bk_0+\de\bn)-\la_0(\bk_0)= C_0\de^{\al}+O(\de^{\al+1}) 
\ene
as $\de\to 0$ with even $\al\ge 4$ and $C_0>0$. We denote $\bn^{\perp}$ to be any of the two unit vectors orthogonal to $\bn$.   
Let $\bnu$ be a vector that belongs to $\tilde\Gad$ with some choice of sufficiently large $N$, but does not belong to $\Gad/2$ (so that $\bka_0+\bnu$ and $\bka_0-\bnu$ are different points modulo $\Gad$). We denote  $l_0:=L(\bk_0)$ so that $\bk_0=\bka_0+\bp_{l_0}$. Denote $v:=\be_{\bnu}+\be_{-\bnu}$  and  
$
H_{\ep}:=H+\ep V,
$
where $V$ is the operator of multiplication by $v$. Whenever $\bka$ 
lies outside of a small neighbourhood of $\bka_0:=\kappa(\bk_0)$, all eigenvalues of $H_{\ep}(\bka)$ are located far away from $\mu_{+}$, therefore we are interested only in quasimomenta $\bka$ located in an $o(1)$-neighbourhood of $\bka_0$ as $\ep\to 0$. 
Therefore, we have to study the perturbation of the eigenvalues of the fibre operators $H_{\ep}(\bka)$ when $\bka$ is close to $\bka_0$.   
We will write down the action of this operator in the orthonormal basis 
\bee\label{basis}
\{\phi_{j,l}(\bka)\}, \,\, j=0,1,\dots; \,\, l=1,\dots,M.
\ene 
The matrix of $H$ in this basis is diagonal, as we have established above. Let us compute the matrix of $V(\bka)$ (the fibre operator corresponding to $V$ at the point $\bka$). We denote by $l_0^+$ the unique index satisfying 
$\bp_{l_0^+}=\bp_{l_0}+\bnu(\mathrm{mod}\ \tilde\Gad)$; similarly, we define $l_0^-$ by requiring  
$\bp_{l_0^-}=\bp_{l_0}-\bnu(\mathrm{mod}\ \tilde\Gad)$. We will assume, as we can without loss of generality, that in fact we have 
\bee
\bp_{l_0^{\pm}}=\bp_{l_0}\pm\bnu
\ene
(in order to achieve this we need to choose proper representatives $\bp_l$  from corresponding classes of equivalence). 
Then we obviously have 
\bee
V: \be_{\bk}\mapsto \be_{\bk+\bnu}+\be_{\bk-\bnu}. 
\ene
Using \eqref{psi} and \eqref{e}, this implies 
\bee
\bes
V: \psi_j(\bk)\mapsto&
|\To| \sum_{\bth\in\Gad}\hat\psi_j(\bth;\bk)
\sum_{m}{\overline{\hat\psi_m(\bth;\bk+\bnu)}}\psi_m(\bk+\bnu)\\
+&|\To|\sum_{\bth\in\Gad}\hat\psi_j(\bth;\bk)
\sum_{m}\overline{\hat\psi_m(\bth;\bk-\bnu)}\psi_m(\bk-\bnu). 
\end{split}
\ene
This means that if we denote by 
\bee
V_{j_1,l_1; j_2,l_2}(\bka):=\lu V\phi_{j_1,l_1}(\bka),\phi_{j_2,l_2}(\bka)\ru_{L^2(\tilde\To)} 
\ene
the matrix element of operator $V$ in the basis \eqref{basis} then we have:
\bee\label{matrix}
V_{0,l_0; m,l}(\bka)=\begin{cases}
|\To|\sum_{\bth\in\Gad}\hat\psi_0(\bth;\bka+\bp_{l_0})
\overline{\hat\psi_m(\bth;\bka+\bp_{l_0^{\pm}})},& {\textrm {if}} \ l=l_0^{\pm}\\
0, & {\textrm {otherwise}}. 
\end{cases}
\ene
Suppose, $\bka$ is close to $\bka_0$ so that the eigenvalue $\la_0(\bka+\bp_{l_0})$ is a simple eigenvalue of $H(\bka)$. Then for small $\ep$ the operator 
$H_{\ep}(\bka)$ will have a single eigenvalue $\tau=\tau_{\ep}(\bka)$ which is inside $2\ep$-neighbourhood of $\la_0(\bka)$.  
Let us write a perturbation theory expansion of $\tau$. Let $P_0$ be the projection onto $\phi_{0,l_0}$ and $P':=I-P_0$. Using $P_0VP_0=0$ (see Corollary \ref{cor:1} or \eqref{matrix}) we obtain:
\bee\label{3.11}
\tau_{\ep}(\bka+\bp_{l_0})=\la_0(\bka+\bp_{l_0})
+Z\ep^2+Y(\ep) 
\ene 
as $\ep\to 0$. Here, 
\bee
Z:=-Tr P_0 VP'(H_0(\bka+\bp_{l_0})-\lambda_0(\bka+\bp_{l_0}))^{-1}P'VP_0 
\ene
and $Y(\ep)=O(\ep^3)$ is analytic in $\bka$ in some neighbourhood of $\bka_0$.   
Next, let $P_1$ be the orthogonal projection onto the two-dimensional subspace generated by $\phi_{0,l_0^\pm}$; we also put $P_1':=P'-P_1$. Then we can rewrite $Z$ as follows.
\bee\label{Z'}
Z=-Tr P_0 VP_1(H_0(\bka+\bp_{l_0})-\lambda_0(\bka+\bp_{l_0}))^{-1}P_1VP_0+R_0,
\ene
where $R_0:=-Tr P_0 VP_1'(H_0(\bka+\bp_{l_0})-\lambda_0(\bka+\bp_{l_0}))^{-1}P_1'VP_0$ is analytic in $\bka$ in sufficiently small neighbourhood of $\bka_0$ (see also \eqref{S} below). Using \eqref{matrix} we get
\bee\label{Z}
Z=\sum_{\pm}\frac{|\ |\To|\sum_{\bth\in\Gad}\hat\psi_0(\bth;\bk)
\overline{\hat\psi_0(\bth;\bk\pm\bnu)}|^2}
{\la_0(\bk)-\la_0(\bk\pm\bnu)}+R_0,
\ene
and 
\bee\label{S}
R_0=\sum_{m\not=0}\frac{|\ |\To|\sum_{\bth\in\Gad}\hat\psi_0(\bth;\bk)
\overline{\hat\psi_m(\bth;\bk+\bnu)}|^2}
{\la_0(\bk)-\la_m(\bk+\bnu)}+\sum_{m\not=0}\frac{|\ |\To|\sum_{\bth\in\Gad}\hat\psi_0(\bth;\bk)
\overline{\hat\psi_m(\bth;\bk-\bnu)}|^2}
{\la_0(\bk)-\la_m(\bk-\bnu)}.
\ene
Note that expressions \eqref{Z} and \eqref{S} depend only on $\bk$ and $\bnu$ and do not depend on $N$ and $\tilde\Gad$. 

\bel\label{lem:Z}
There exists a vector $\bnu$ of arbitrarily small positive length such that
\bee\label{nonzero}
\frac{\partial^2Z}{\partial \bn^2}(\bk_0,\bnu)\ne 0. 
\ene
\enl
\bep
We will consider vectors $\bnu:=\de \bn$ with small $\de$ and write the expansion of $ \frac{\partial^2Z}{\partial \bn^2}(\bk_0,\bnu)$ in terms of $\de$ when
$\de\to 0$. We will transform the formula for $Z$ in the following way: 
\bee\label{Z1}
\bes
Z&=\frac{|\ |\To|\sum_{\bth\in\Gad}\hat\psi_0(\bth;\bk)
\overline{\hat\psi_0(\bth;\bk)}|^2}
{\la_0(\bk)-\la_0(\bk+\bnu)}+\frac{|\ |\To|\sum_{\bth\in\Gad}\hat\psi_0(\bth;\bk)
\overline{\hat\psi_0(\bth;\bk)}|^2}
{\la_0(\bk)-\la_0(\bk-\bnu)}+R+R_0\\
&=\frac{1}{\la_0(\bk)-\la_0(\bk+\bnu)}+\frac{1}{\la_0(\bk)-\la_0(\bk-\bnu)}+R+R_0.
\end{split}
\ene

The second equality in
\eqref{Z1} is due to \eqref{psi1}. 
Note that $\la_0(\bk\pm\bnu)$, $\hat\psi_0(\bth;\bk\pm\bnu)$, and $R_0$ are analytic in $\de$ near $0$ and in $\bk$ near $\bk_0$. 
It is a straightforward calculation that $R=R(\de)=R(\de;\bk,\bnu)$ satisfies 
$R(\de)=O(\de^{-\al+1})$, $R'(\de)=O(\de^{-\al})$, and $\ R''(\de)=O(\de^{-\al-1})$ as  $\de\to 0$ (compare with calculations \eqref{nn1} and \eqref{nn2} and the proof of Lemma~\ref{remark} below; recall that $\alpha$ is defined in \eqref{alpha}).  Let us calculate now the second derivative of the principle term in \eqref{Z1}. We have (here, by $f'$ we denote $\frac{\partial f}{\partial \bn}$):
\bee\label{nn1}
\bes
&\frac{\partial^2}{\partial \bn^2}\left( \frac{1}{\la_0(\bk)-\la_0(\bk+\bnu)}\right)\\ 
=& - \frac{\partial}{\partial \bn}\left( \frac{\la_0'(\bk)-\la_0'(\bk+\bnu)}{(\la_0(\bk)-\la_0(\bk+\bnu))^2}\right)\\
&=2\frac{(\la_0'(\bk)-\la_0'(\bk+\bnu))^2}{(\la_0(\bk)-\la_0(\bk+\bnu))^3}-\frac{\la_0''(\bk)-\la_0''(\bk+\bnu)}{(\la_0(\bk)-\la_0(\bk+\bnu))^2}. 
\end{split}
\ene
Therefore, 
\bee\label{nn2}
\bes
&\frac{\partial^2}{\partial \bn^2}\left( \frac{1}{\la_0(\bk)-\la_0(\bk+\bnu)}\right)\Bigm|_{\bk=\bk_0}\\
=&
\frac{2\la_0'(\bk_0+\bnu)^2+\la_0''(\bk+\bnu)(\la_0(\bk)-\la_0(\bk+\bnu))}{(\la_0(\bk)-\la_0(\bk+\bnu))^3}\\
\sim & \frac{ 2C_0^2\al^2\de^{2\al-2} - \al(\al-1)C_0^2 \de^{2\al-2} }{-C_0^3\de^{3\al}}=
-C_0^{-1}\al(\al+1)\de^{-\al-2}. 
\end{split}
\ene
It is easy to check that the contribution from the derivative of the second term in \eqref{Z1} is the same. Therefore, as $\de\to 0$, we have 
\bee
 \frac{\partial^2Z}{\partial \bn^2}(\bk_0,\bnu)=-2C_0^{-1}\al(\al+1)\de^{-\al-2}+o(\de^{-\al-2}),
 \ene
which shows that for some $\bnu$ this second derivative is non-zero. 
\enp

Now, we are going to make the formula for $Z$ more specific by introducing the proper coordinates. 

\bel \label{remark}
Let us denote the coordinates of $\bk$ around $\bk_0$ by $(x,y)$ so that $x:=\lu\bk-\bk_0,\bn\ru$ is the coordinate along $\bn$, and $y:=\lu(\bk-\bk_0),\bn^{\perp}\ru$ is the coordinate along $\bn^{\perp}$. Then we have: 
\bee\label{Zfull}
Z=O(\delta^{-\al})+O(\delta^{-\al-1})x-C_1\delta^{-\al-2}(1+O(\delta))x^2+C(\delta)x^3+
yf(x,y),\ C_1>0,
\ene
where $f$ is a real analytic function in a neighbourhood of the origin. Function $f(x,y)$ also depends on $\delta$ but actual dependence is not important. 
\enl

\bep
In our new coordinates formula \eqref{Z1} looks as follows
$$
Z=\frac{1}{\la_0(x,y)-\la_0(x+\delta,y)}+\frac{1}{\la_0(x,y)-\la_0(x-\delta,y)}+R+R_0,
$$
where $R_0$ is analytic in $(x,y,\delta)$ near zero and 
\bee
\begin{split}& 
R=\frac{|\ |\To|\sum_{\bth\in\Gad}\hat\psi_0(\bth;x,y)
\overline{\hat\psi_0(\bth;x+\delta,y)}|^2-|\ |\To|\sum_{\bth\in\Gad}\hat\psi_0(\bth;x,y)
\overline{\hat\psi_0(\bth;x,y)}|^2}
{\la_0(x,y)-\la_0(x+\delta,y)}\cr &+\frac{|\ |\To|\sum_{\bth\in\Gad}\hat\psi_0(\bth;x,y)
\overline{\hat\psi_0(\bth;x-\delta,y)}|^2-|\ |\To|\sum_{\bth\in\Gad}\hat\psi_0(\bth;x,y)
\overline{\hat\psi_0(\bth;x,y)}|^2}
{\la_0(x,y)-\la_0(x-\delta,y)}.
\end{split}
\ene
We recall that $\hat\psi_0$ and $\lambda_0$ are real-analytic in $(x,y)$ near zero. Equation \eqref{alpha} has the form
$$
\lambda_0(\delta,0)=C_0\delta^\alpha+O(\delta^{\alpha+1})
$$
or, in other words,
$$
\lambda_0(x,y)=C_0 x^\alpha+O(x^{\alpha+1})+O(y).
$$
Now, the direct calculations similar to the ones from the proof of Lemma~\ref{lem:Z} complete the proof of \eqref{Zfull} which is just the Taylor series for $Z$ in $x$ and $y$ near zero with $\delta$ being a parameter. We emphasize that while $\delta$ is going to be small (for example, to ensure that the coefficient in front of $x^2$ is negative and the eigenvalue $\lambda_0$ is still simple after the shift of the argument by $\delta$) it will be fixed; then $x$ and $y$ are considered to be in a small neighborhood (depending on $\delta$) of the origin which ensures the convergence of the Taylor series \eqref{Zfull}.
\enp

\ber
Since formula \eqref{Zfull} is an identity of analytic in $x,y$ functions (for every sufficiently small $\de>0$), we can differentiate this identity with respect to $x$ and $y$ arbitrary many times. 
\enr

Now we discuss the broad strategy of our approach. Suppose, $H=:H^0$ is our initial 
operator with a degenerate minimum of the spectral edge (meaning that  $S$ consists of several isolated points, but the quadratic form of the Bloch function $\la(\bk)$ at one of them is degenerate). We then start perturbing $H$ by adding potentials of the form $\ep_j(\be_{\bnu_j}+\be_{-\bnu_j})$, so that  $H^{j}=H^{j-1}+\ep_j(\be_{\bnu_j}+\be_{-\bnu_j})$. Each $H^{j}$ is a periodic operator with the lattice of periods $\Ga_j$, where each $\Ga_j$ is a sub-lattice of $\Ga_{j-1}$. At each step we will achieve that a certain partial derivative (or a certain combination of partial derivatives) of a perturbed Bloch function at the new extremal point (or points) becomes non-zero. 
Lemma \ref{continuity} shows that, 
once some combination of the partial derivatives on the Bloch function is non-zero for the operator $H^j$, we can choose $\ep_n$, $n>j$ so small that the same combination is non-zero for all operators $H^n$, $n\ge j$ (notice that the choice of how small we require each $\ep_n$ to be depends also on the lattice $\Gamma_n$). At the end our objective is to achieve that at all local minima of the Bloch functions $\la(\bka)$ of $H^n$ located near $\mu_{+}$  we had $\partial_{xx}\la\ne 0$, $\partial_{yy}\la\ne 0$, and $\partial_{xx}\la\partial_{yy}\la-(\partial_{xy}\la)^2\ne 0$ in some coordinate system $(x,y)$; then, all minima will be non-degenerate. Of course, we will also make sure that $\ep_j$ are so small that $\sum_j\ep_j<\mu_{+}-\mu_{-}$, so we have not closed the spectral gap. We will also assume that the perturbed operator is still periodic with, possibly, a
new lattice of periods $\tilde\Ga\subset\Ga$. In order to achieve this, it is enough to require that each vector $\bnu_j$ belongs to the set $\Q\Gad$. This set is dense and the objective of our perturbation will always be making certain quantities (like combinations of partial derivatives) non-zero. Since these quantities will always depend continuously on $\bnu$, once we have found any vector $\bnu$ for which these quantities are non-zero, we can always find a vector inside  $\Q\Gad$ with these quantities still being non-zero. For example, in Lemma \ref{lem:Z} we can always find $\bnu$ satisfying the requirements of that Lemma such that, additionally, we have $\bnu\in \Q\Gad$. 
Therefore, we will always assume that our choice of $\bnu_j$ will be rational multiples of a vector from $\Ga$, without specifying it explicitly. 

\subsection{Several band functions have the same minimum at the same point}
In this section, we will get rid of a situation when a point $\bk_0$ is a minimum of two or more different band functions simultaneously. It was proved in \cite{KlRa} that (generically) this cannot happen; here, we will give (an outline of) a different proof, which seems to us to be rather shorter. 
We will need the following Lemma, sometimes known as the Shur complement Lemma. 

\bel\label{lem:1}
Suppose, $P_1$ and $P_2$ are two orthogonal projections in a Hilbert space $\GH$ with $P_1+P_2=I$, and $H$ is the self-adjoint operator which has the following block form with respect to $P_1$ and $P_2$. 
\bee
H=\left(
\begin{array}{cc}
U_{11} & U_{12} \\
U_{21} & U_{22} 
\end{array}
\right). 
\ene
This means that $U_{jl}=P_jHP_l$. We put $\GH_j:=P_j(\GH)$ and assume that $\la\not\in\sigma(U_{22})$. 

1. Suppose, $\psi\in \GH_1$ is a vector lying in the kernel of $(U_{11}-\la)-U_{12}(U_{22}-\la)^{-1}U_{21}$. Then 
$\tilde\phi:= \left(
\begin{array}{c}
\psi \\
\phi 
\end{array}
\right)$, where $\phi:=-(U_{22}-\la)^{-1}U_{21}\psi$, is an eigenvector of $H$ corresponding to $\la$. 

2. Suppose, $\la$ is an eigenvalue of $H$. Then the kernel of $(U_{11}-\la)-U_{12}(U_{22}-\la)^{-1}U_{21}$ (considered as an operator in $\GH_1$) is non-trivial.   
\enl

\bep
This is a straightforward computation. 
\enp

To begin with, let us assume that $\mu_{+}$ is the minimum of two band functions reached at the same point, say $\la_1(\bk_0)=\la_2(\bk_0)=\mu_{+}$. In this case we will not be 
taking vector $\bnu$ from a finer lattice $\tilde\Gad$, instead we will assume that $\bnu\in\Gad\setminus\{0\}$. We also take slightly more complicated potential than before, namely we put $v=a\be_{\bnu}+\bar{a}\be_{-\bnu}$ and denote by $V$ the operator of multiplication by $v$. 
Let us check what will happen with the eigenvalues at $\bk_0$ after this perturbation. We apply the Shur complement Lemma to study eigenvalues of $H_{\ep}(\bk)$ with $\bk$ close to $\bk_0$. We denote by $P_1$ the  orthogonal projection onto $\GH(\bk)$ -- the two-dimensional subspace generated by $\psi_1(\bk;\cdot)$ and $\psi_2(\bk;\cdot)$ and $P_2=I-P_1$. 
Then the Shur complement Lemma shows that the perturbed eigenvalues coincide with the eigenvalues of the $2\times 2$ matrix $A=A(\varepsilon,\bk)=(a_{mn})_{m,n=1}^2$ with the coefficients given by  
\bee
\bes
&a_{mm}
=\la_m(\bk)\\
&+\ep[a\sum_{\bth\in\Gad}\hat\psi_m(\bth;\bk)
\overline{\hat\psi_m(\bth;\bk+\bnu)}+\bar{a}\sum_{\bth\in\Gad}\hat\psi_m(\bth;\bk)
\overline{\hat\psi_m(\bth;\bk-\bnu)}]+O(\ep^2)
\end{split}
\ene
and 
\bee
a_{mn}
=\ep[a\sum_{\bth\in\Gad}\hat\psi_m(\bth;\bk)
\overline{\hat\psi_n(\bth;\bk+\bnu)}+\bar{a}\sum_{\bth\in\Gad}\hat\psi_m(\bth;\bk)
\overline{\hat\psi_n(\bth;\bk-\bnu)}]+O(\ep^2)
\ene
if $m\ne n$. 
We notice that the choice of the basis in $\GH(\bk_0)$ is not uniquely determined; we just fix some orthonormal basis  $\phi:=(\psi_1(\bk_0;\cdot),\psi_2(\bk_0;\cdot))$ of $\GH(\bk_0)$.
\bel
For some $a\in\C$ and $\bnu\in\Gad\setminus\{0\}$ we have 
\bee\label{offdiag}
a\sum_{\bth\in\Gad}\hat\psi_1(\bth;\bk_0)
\overline{\hat\psi_2(\bth;\bk_0+\bnu)}+\bar{a}\sum_{\bth\in\Gad}\hat\psi_1(\bth;\bk_0)
\overline{\hat\psi_2(\bth;\bk_0-\bnu)}\ne 0.
\ene
\enl
\bep
Suppose not. Then for each $\bnu\in\Gad$ we have $\sum_{\bth\in\Gad}\hat\psi_1(\bth;\bk_0)
\overline{\hat\psi_2(\bth;\bk_0+\bnu)}=0$ (this sum is zero if $\bnu=0$ due to \eqref{psi1} anyway). Notice that 
\bee
\sum_{\bth\in\Gad}\hat\psi_1(\bth;\bk_0)
\overline{\hat\psi_2(\bth;\bk_0+\bnu)}=
\sum_{\bth\in\Gad}\hat\psi_1(\bth;\bk_0)
\overline{\hat\psi_2(\bth+\bnu;\bk_0)},
\ene
and these numbers are Fourier coefficients of the product 
$\psi_1(\bk_0,\bx)\overline{\psi_2(\bk_0,\bx)}$. This product, however, cannot be identically equal to zero for all $\bx$ due to the unique continuation.
\enp

This lemma shows that off-diagonal elements of $A(\varepsilon,\bk_0)$ are non-zero for a certain choice of 
$a$ and $\bnu\ne 0$. Therefore, its eigenvalues are different and we have achieved the required splitting. This simple argument is already sufficient to prove that Condition A is generic if eigenfunctions $\psi_j$ are continuous in $\bk$, since then off-diagonal elements of $A$ will be non-zero for all $\bk$ in a neighbourhood of $\bk_0$. 

In general situation we proceed as follows. We denote by $\{v_{mn}(\bk)\}_{m,n=1}^2$ the matrix of $V$ in the basis $\psi_j(\bk)$, $j=1,2$, and 
for any choice $\phi=(\phi_1,\phi_2)$  of an orthonormal basis of $\GH(\bk_0)$  we denote 
by $\{v_{mn}(\bk_0;\phi)\}_{m,n=1}^2$ the matrix of operator $V$ in this basis. 
Let us fix $a$ and $\bnu$ so that the left-hand side of \eqref{offdiag} is equal to one, i.e. $v_{12}(\bk_0;\psi)=1$. 
This means that for any other choice of the basis $\phi$ we either have $|v_{12}(\bk_0;\phi)|\geq 1/4$ or $|v_{11}(\bk_0;\phi)-v_{22}(\bk_0;\phi)|\geq 1$. Since the projection onto 
$\GH(\bk)$ is analytic in $\bk$, for every $\eta>0$ there exists $\delta>0$ such that whenever $|\bk-\bk_0|\leq\delta$ the matrix $\{v_{mn}(\bk)\}_{m,n=1}^2$ 
is $\eta$-close to the matrix $\{v_{mn}(\bk_0;\phi)\}_{m,n=1}^2$ with {\it some} choice of the basis $\phi$. 
As a consequence, taking $\eta=1/8$ we obtain the following statement. For every $\bk$ in some neighbourhood $M$ of $\bk_0$ the elements of the matrix $\{v_{mn}(\bk)\}_{m,n=1}^2$ satisfy either $|v_{12}(\bk)|\geq 1/8$ or $|v_{11}(\bk)-v_{22}(\bk)|\geq 3/4$.

Now, we consider the matrix of $A(\varepsilon,\bk)$ in the basis $\psi_j(\bk)$, $j=1,2$. We have 
$$a_{jj}(\bk)=\lambda_j(\bk)+\varepsilon v_{jj}(\bk)+O(\varepsilon^2),\ \ a_{12}(\bk)=\varepsilon v_{12}(\bk)+O(\varepsilon^2).$$
Here, $O(\varepsilon^2)$ is uniform in $\bk\in M$. If $\bk$ is such that $|v_{12}(\bk)|\geq 1/8$ then we obviously don't have a multiple eigenvalue for $\varepsilon$ small enough to dominate $O(\varepsilon^2)$. Assume the second alternative, i.e. $|v_{12}(\bk)|\leq 1/8$ but $|v_{11}(\bk)-v_{22}(\bk)|\geq 3/4$. Assume for definiteness $v_{11}(\bk)>v_{22}(\bk)+3/4$ and $\lambda_1(\bk_0)=\lambda_2(\bk_0)=0$, and $0$ is the minimal value of these functions. We also have $|v_{jj}(\bk)-v_{jj}(\bk_0;\phi)|\leq 1/8$ for some choice of the basis $\phi$. The matrix of $A(\varepsilon,\bk_0)$ in this basis has the form
$$a_{jj}(\bk_0;\phi)=\varepsilon v_{jj}(\bk_0;\phi)+O(\varepsilon^2),\ \ a_{12}(\bk_0;\phi)=\varepsilon v_{12}(\bk_0;\phi)+O(\varepsilon^2).$$
Now, assume that we have a multiple eigenvalue at $\bk\in M$. Then $a_{22}(\bk)=a_{11}(\bk)$, which, together with the observation  that $\lambda_j(\bk)\geq0$, implies  $\lambda_2(\bk)\geq 3\varepsilon/4 +O(\varepsilon^2)$ and thus 
$$
a_{22}(\bk)\geq \varepsilon(3/4+v_{22}(\bk))+O(\varepsilon^2)\geq \varepsilon(5/8+v_{22}(\bk_0;\phi))+O(\varepsilon^2)> a_{22}(\bk_0;\phi).
$$
this means that $\bk$ is not a point of minimum. Thus we proved that in some neighbourhood of $\bk_0$ we cannot have the edge of the spectrum attained by more than one band function.

Suppose now that  $\mu_{+}$ is the minimum of $t$ band functions reached at the same point, $\la_1(\bk_0)=\la_2(\bk_0)=\dots=\la_t(\bk_0)=\mu_{+}$. Then we proceed as above and our perturbation will be described by a $t\times t$ matrix the off-diagonal elements of which are non-zero for some choice of parameters. Arguments similar to those above imply that  after this perturbation, the resulting operator will have an eigenvalue of multiplicity at most $t-1$. Repeating this procedure $t-1$ times if necessary, we will achieve that no two different bands can have a minimum at the same point. The arguments which justify the subsequent elimination of the multiple minima at different points or/and at different edges are standard. 
This proves the following Theorem (originally due to Klopp-Ralston, \cite{KlRa}): 

\bet
Condition A is generic for two-dimensional periodic potentials.
\ent

\subsection{Minimum is a minimum of only one band function}

First, we choose coordinates around $\bk_0$ so that $x$ goes along $\bnu$. 
In this section we will change these coordinates many times; in order to avoid cumbersome notation, we will call both old and new set of coordinates by the same letters $(x,y)$ (sometimes writing $x_{old}$ and $x_{new}$ to avoid confusion). Each time we perform a change of coordinates, we will have to check that the perturbation $Z$ in the new coordinates still satisfies \eqref{Zfull} (or, at least, \eqref{nonzero}). 


Step 1. Obtaining a quadratic term in one direction. 

Let $\bk_0$ be a point of local minimum of $\la_0$; we will introduce the orthogonal coordinates $(x,y)$ around $\bk_0$ so that the Taylor expansion of $\la_0$ 
at $\bk_0$ in these coordinates has a form 
\bee
\la_0(x,y)=x^{2n}+\sum_{\al,\beta}d_{\al\beta}x^{\al}y^{\beta}
\ene
and the sum is over all $(\al,\beta)$ with $\al+\beta\ge 2n$ with the exception of $(\al,\beta)=(2n,0)$. If $n=1$, we move to the next step, so now we assume that $n\ge 2$. 
We apply the Weierstrass Preparation Theorem and obtain 
that $\la_0$ has the following form:
$$
\lambda_0(x,y)=(x^{2n}+\sum\limits_{j=0}^{2n-1}a_j(y)x^j)c(x,y),
$$
where $a_j$ are analytic functions such that $a_j(y)=O(y^{2n-j})$ and $c(0,0)=1$. Making a change of variables $x_{new}=x_{old}-a_{2n-1}(y)/(2n)$, we can assume that $a_{2n-1}=0$. We notice that this  change of variables does not affect the representation $\eqref{Zfull}$. So, for simplicity we will use the same notation $(x,y)$ for the new variables. Then the Bloch function after the perturbation has a form 
\bee
\bes
&\tau_{\ep}(x,y)=\lambda_0(x,y)+\varepsilon^2Z(x,y)+\varepsilon^3b(x,y,\ep)\\
&=c(x,y)\left(x^{2n}+\sum\limits_{j=0}^{2n-2}a_j(y)x^j+
\varepsilon^2Z(x,y)/c(x,y)+\varepsilon^3b(x,y,\ep)/c(x,y)\right).
\end{split}
\ene
Here, $b$ is analytic function in all variables. 
Each $a_j$ is analytic function of one variable and therefore has a simple form $a_j(y)=c_jy^{k_j}(1+O(y)),\ c_j\not=0$ (either this, or $a_j\equiv 0$, in which case we put $k_j:=\infty$).  Obviously, $k_j\geq 2n-j$. 
Let $(x_*,y_*)$ be a point where the minimum of $\tau_{\ep}$ is attained in a small neighbourhood of the origin, i.e. $x_*,\,y_*=o(1)$ as $\ep\to 0$. We prove
that there is an improvement after the perturbation, namely:
\bel\label{step1}
There is a partial derivative of $\tau_{\ep}$ of order smaller than $2n$ that does not vanish at $(x_*,y_*)$. 
\enl
\bep
Assume that it is not so. Then all partial derivatives of $\tau_{\varepsilon}$ of order smaller than $2n$ are equal to zero at  $(x_*,y_*)$. It is easy to see that $\tau_\ep(x_*,y_*)=O(\varepsilon^2)$. Indeed, in any case $\tau_\ep(0,0)=O(\varepsilon^2)$ and $\tau_\ep(x_*,y_*)\geq -C\varepsilon^2$ (as a sum of a non-negative function $\lambda_0$ and $O(\varepsilon^2)$). Thus, either $\tau_\ep(x_*,y_*)=O(\varepsilon^2)$, or $\tau_\ep(x_*,y_*)>\tau_\ep(0,0)$, in which case $(x_*,y_*)$ cannot possibly be a minimum of $\tau$. 

Thus, all partial derivatives of $\tilde\tau:=\tau_{\varepsilon}/c(x,y)$ of order smaller than $2n$ are $O(\varepsilon^2)$ at  $(x_*,y_*)$. We get
$$
O(\varepsilon^2)=\frac{\partial^{2n-1}\tilde\tau}{\partial x^{2n-1}}=(2n)!x_*+O(\varepsilon^2),
$$
which gives $x_*=O(\varepsilon^2)$. Next, for $j=0,\dots,2n-2$,
$$
O(\varepsilon^2)=\frac{\partial^{2n-1}\tilde\tau}{\partial x^j\partial y^{2n-1-j}}=c_j\frac{j!k_j!}{(k_j-2n+1+j)!}y_*^{k_j-2n+1+j}(1+o(1))+O(\varepsilon^2),
$$
and this implies $y_*=O(\varepsilon^{2/(k_j-2n+1+j)})$ for $j=0,\dots,2n-2$.

Now we notice that formula \eqref{Zfull} implies that 
\bee
Z(x,y)=b_0+b_1x-b_2x^2+O(x^3)+O(y),
\ene
where $b_j$ are functions of $\de$ and $b_2>0$ for small $\de$. Then our assumption that all derivatives of $\tau_{\varepsilon}$ of order smaller than $2n$ disappear implies 
\begin{equation*}
\bes
& 0=\frac{\partial^{2}\tau_\ep}{\partial x^{2}}\cr &=2c_2y_*^{k_2}(1+o(1))c(x_*,y_*)+2c_1y_*^{k_1}(1+o(1))c'_x(x_*,y_*)+c_0y_*^{k_0}(1+o(1))c''_{xx}(x_*,y_*)\cr &-2b_2\varepsilon^2+o(\varepsilon^2),
\end{split}
\end{equation*}
assuming $k_j$, $j=0,1,2$, are finite. 
But since $k_j>k_j-2n+1+j,\ j=0,1,2,$ for $2n\geq4$, this is the contradiction.
The case when one or more $k_j=\infty$ is even simpler and can be considered in the same way. 
\enp
This lemma shows that after the perturbation we get a non-zero derivative of order smaller than $2n$ in the new (and therefore in the old) variables. Repeating this procedure, we obtain a new Bloch function for which the second derivative in one direction does not vanish. Now we move to the next step. 

Step 2. Obtaining a nondegenerate quadratic form.

Suppose now that the second derivative at our minimum in one direction is non-degenerate, i.e. 
\bee\label{degenerate}
(\lambda_0)''_{yy}(0,0)\not=0.
\ene
If second derivatives in all directions are non-degenerate, we have nothing else to do, so we also assume that
\bee\label{degenerate22}
(\lambda_0)''_{xx}(0,0)=0.
\ene
By Weisstrass Preparation Theorem $\lambda_0=(y^2+2f_1(x)y+f_2(x))p(x,y)$ with some analytic functions $f_1$, $f_2$, $p$, such that $p(0,0)>0$. 
We immediately notice that $f_1(x)=O(x^2)$. Indeed, if $f_1(x)$ has a non-trivial linear term in its Taylor expansion, then, since $(0,0)$ is a minimum, $f_2(x)$ must have a non-trivial quadratic term, which contradicts \eqref{degenerate22}. 
Changing variables $x_{new}= x_{old},\ y_{new}= y_{old}+f_1(x_{old})$ we get 
$$
\lambda_0=(y^2 + f(x)) p_1(x,y),\ \ \ p_1(0,0)>0.
$$
Since we have an isolated minimum, $f(x)=bx^{2n}(1+O(x))$ with some $b>0$ and $n\geq 2$. Now, rescaling we obtain that $\lambda_0$ has the form $\lambda_0=(x^{2n}(1+O(x))+y^2)c(x,y)$ with $c(0,0)>0$. Finally, we make another change of variables $ x_{new}^{2n}=x_{old}^{2n}(1+O(x_{old}))c(x_{old},y_{old})$ and $y_{new}^2=y_{old}^2c(x_{old},y_{old})$ so that in the new coordinates we have 
\bee
\tilde\tau=x^{2n}+y^2+\ep^2 Z+O(\ep^3),  
\ene
where we have denoted $\tilde\tau(x_{new},y_{new})=\tau(x_{old},y_{old})$. 
 Below (see Lemma~\ref{Znew}) we will show that $Z$ from \eqref{Zfull} still admits similar representation in new variables:
\begin{equation}\label{ZZnew} Z(x,y)=b_0+b_1x-b_2x^2+O(x^3)+O(y),\ \ \ b_2>0.\end{equation}
As before, we assume that $(x_*,y_*)$ is a point of a local minimum for $\tilde\tau_\varepsilon$ near point $(0,0)$ (in particular, $x_*$, $y_*=o(1)$). 
We consider three cases. 

Case 1. Suppose, $b_1\not=0$. Then from $\nabla\tilde\tau(x_*,y_*)=0$ we get $y_*=O(\varepsilon^2)$, $x_*=\left(\frac{-b_1}{2n}\varepsilon^2\right)^{\frac{1}{2n-1}}(1+o(1))$. This implies 
$$
(\tilde\tau)''_{yy}(x_*,y_*)=2+O(\varepsilon^2),\ \ (\tilde\tau)''_{xy}(x_*,y_*)=O(\varepsilon^2),$$$$(\tilde\tau)''_{xx}(x_*,y_*)=2n(2n-1)\left(\frac{-b_1}{2n}\varepsilon^2\right)^{\frac{2n-2}{2n-1}}(1+o(1))+O(\varepsilon^2).
$$
Thus, we have a nondegenerate minimum at $(x_*,y_*)$. 

Case 2. Suppose, $b_1=0$ and $n\geq3$. We notice that if $x_*=O(\varepsilon)$ then $(\tilde\tau)''_{xx}(x_*,y_*)<0$ which leads to a contradiction. So, $|x_*/\varepsilon|\to\infty$ as $\varepsilon\to0$. Then, similar to the previous case one gets $y_*=O(\varepsilon^2)$, $x_*=\left(\frac{b_2}{n}\varepsilon^2\right)^{\frac{1}{2n-2}}(1+o(1))$ and
$$
(\tilde\tau)''_{yy}(x_*,y_*)=2+O(\varepsilon^2),\ \ (\tilde\tau)''_{xy}(x_*,y_*)=O(\varepsilon^2),$$$$(\tilde\tau)''_{xx}(x_*,y_*)=2n(2n-1)\left(\frac{b_2}{n}\varepsilon^2\right)(1+o(1))-2b_2\varepsilon^2.
$$
Thus, we again have nondegenerate minimum at $(x_*,y_*)$.

Case 3. Finally, we consider the case $b_1=0$ and $n=2$. It is convenient to rescale $x_{old}=\varepsilon x_{new}$, $y_{old}=\varepsilon^2y_{new}$ and divide by $\varepsilon^4$. Then we have to consider
\begin{equation}\label{mu1}
\hat\tau=x^4+y^2+O_{x,y}(1)\varepsilon-b_2 x^2+O(1)x+O(1)y+O(1)\varepsilon^{-2}.
\end{equation}
Here we have used the fact that $Y(\varepsilon)$ from \eqref{3.11} is analytic in $(x,y)$. We are using the following convention: $O(1)$ is a bounded function of $\varepsilon$ only, and $O_{x,y}(1)$ is a bounded analytic function of $\varepsilon, x,y$.  
Calculating the derivatives at point $(x_*,y_*)$ we get
\begin{equation}\label{mu3}
\begin{split}&
\hat\tau=(1+O(\varepsilon))(x-x_*)^4 + (1+O(\varepsilon))(y-y_*)^2+\cr & (4x_*+O(\varepsilon))(x-x_*)^3+(6x_*^2-b_2+O(\varepsilon))(x-x_*)^2+const+O_{x,y}(1)\varepsilon.
\end{split}
\end{equation}

Since we have a minimum, the worst scenario is when our quadratic form is degenerate. This means that
$$
6x_*^2-b_2=O(\varepsilon)
$$
and corresponding form becomes zero in the direction $(y-y_*)=O(\varepsilon)(x-x_*)$. In this direction the cubic term becomes $(4x_*+O(\varepsilon))(x-x_*)^3$ which contradicts to minimum condition since $x_*^2\sim b_2/6$.

Now, let us show that the change of variables we use above does not destroy our achievements, i.e. that \eqref{ZZnew} holds in the new variables.
\bel\label{Znew}
All changes of the variables described above do not change the representation \eqref{Zfull} for sufficiently small $\de$. In particular, \eqref{ZZnew} holds.
\enl
\begin{proof}
First, we discuss the change of variables 
\bee\label{change1}
x_{new}=x_{old},\ \ \  y_{new}=y_{old}+f_1(x_{old}). 
\ene
If $f_1(x)=O(x^3)$ then the statement immediately follows from \eqref{Zfull}. So, we assume that $f_1(x)=sx^2(1+O(x)),\ s\not=0$. Then, since we have a minimum at point $(0,0)$ we have 
$f_2(x)=as^2x^4(1+O(x)),\ a\geq1$. 
Using this explicit form for $\lambda_0=(y^2+2f_1(x)y+f_2(x))p(x,y)$ and repeating the calculations similar to \eqref{nn1}, \eqref{nn2},  it is not difficult to obtain more detailed version of \eqref{Zfull}. Namely,
\bee\label{Zfull1}
\begin{split}& 
Zp(0,0)=O(\delta^{-4})+O(\delta^{-5})x-\frac{20\delta^{-6}}{as^2}(1+O(\delta))x^2+\frac{4\delta^{-6}}{a^2s^3}(1+O(\delta))y+O(x^3)+O(y^2)\cr & 
=O(\delta^{-4})+O(\delta^{-5})\tilde x-\left(\frac{20}{as^2}+\frac{4}{a^2s^2}\right)\delta^{-6}(1+O(\delta))\tilde x^2+O(\tilde x^3)+O(\tilde y).
\end{split}
\ene
This proves the statement for the substitution \eqref{change1}. 

Finally, the change of variables of the form $x_{new}=x_{old}(s_1+O(x_{old})+O(y_{old}))$ and $y_{new}=y_{old}(s_2+O(x_{old})+O(y_{old}))$, $s_1s_2\not=0$, does not affect the representation \eqref{Zfull} for sufficiently small $\de$ because the coefficient in front of $x$ have smaller order in $\delta$ than the one in front of $x^2$.
\enp

\subsection{Several minima}\label{choice}
All the results of this Section obtained so far prove our main Theorem under assumptions that on each step of the procedure we have $|S|=1$ (i.e. the minimum of the band function is attained at one point).
Let us discuss the changes we need to make if $S$ consists of several points. Then we have to be slightly more careful with the choice of $\bnu$. The properties we need are summarised in the following statement:

\bel
Suppose $S$ is finite and arbitrarily sufficiently small $\delta>0$ is fixed. Then we can find a vector $\bnu\in\Q\Gad$ arbitrarily close to a given direction with the length $|\bnu|\in (\delta/2,\delta)$ and such that there are no two different points $\bk_1,\bk_2\in S$ satisfying $\bk_1+n\bnu=\bk_2+\bth$, where $n\in\Z$ and $\bth\in\Gad$.
\enl

Let us first discuss why these properties are sufficient for our purposes (and where exactly in our procedure these properties are required). We need to be able to choose $\bnu$ close to any direction to be able to perform Step 2. Here it is important to have uniform control of the length of $\bnu$ so that all the estimates from Step 2 still hold for sufficiently close direction. Since $\delta$ is arbitrarily small we can also ensure that, e.g., $C_1>0$ in \eqref{Zfull}. Let us denote by $\tilde\Gad$ the lattice generated by $\Gad$ and $\bnu$ (this lattice is discrete due to the assumption $\bnu\in\Q\Gad$). Our last assumption means that all points $\bk_j\in S$ are different modulo this new lattice $\tilde\Gad$ (i.e. $\bka_j$ are different). This guarantees that \eqref{3.11} holds. Indeed, without this assumption $Y$ is infinite and with this assumption $Y$, although depending on $\bnu$ in an uncontrolled way, is still $O(\ep^3)$ and analytic in $\bka$ in a neighbourhood of $\bka_0$.  

\bep
We assume that $\delta$ is smaller than $\frac{1}{100}\min\{|\gamma|,\ 0\not=\gamma\in\Gad\}$. Let us start by choosing any vector $\tilde\bmu$ from $\Q\Gad$ with direction close to a given one. 
Let $\bmu$ be the smallest vector in $\Gad$ having the same direction as $\tilde\bmu$. We put $\bnu:=\frac{\tilde p \bmu}{p}$, where $\tilde p$ and $p$ are natural numbers defined as follows. Suppose, $\bk_j,\bk_s\in S$ are two points such that $\bk_j-\bk_s=\frac{m_{js}}{n_{js}}\bmu+\bth
$, where  $n_{js}>1$ and $m_{js}$ are co-prime integers and $\bth\in\Gad$; note that $n_{js}$ is uniquely determined by $\bk_j$ and $\bk_s$ and does not depend on $\bth\in\Gad$. 
If there are no such points $\bk_j,\bk_s$, we just define $\tilde p:=1$ and choose $p$ to be any natural number such that $|\bnu|\in(\delta/2,\delta)$. Otherwise, we first choose $p$ to be any large prime number (namely, $p>100|\bmu|/\delta$) co-prime with all $n_{js}$. Then we choose $\tilde p$ such that $|\bnu|\in(\delta/2,\delta)$. Obviously, $\tilde p$ is smaller than $p$ and thus it is co-prime with $p$. We claim that this choice of $\bnu$ satisfies all the required conditions. Indeed, assume that $\bk_1+n\bnu=\bk_2+\bth$, where $n\in\Z$ and $\bth\in\Gad$. This means that $\frac{m_{12}+qn_{12}}{n_{12}}\bmu+\frac{n\tilde p}{p}\bmu=0$ for integer $n,q$ with $|n|<p$, $n\ne 0$. However, this implies that $n\tilde p n_{12}=-(qn_{12}+m_{12})p$, which is a contradiction since $n_{12}$ and $\tilde p$ are co-prime with $p$, and $0<|n|<p$.

\enp


\section{Counter-examples}\label{example}

First of all, in Subsection \ref{4.1},  we will give several examples of  {\bf discrete} periodic Schr\"odinger operators for which property B is violated on an open set of potentials. This obviously shows that property B cannot possibly be generic in the discrete setting. Then, in Subsection \ref{4.2}, we will discuss how property B can be forced to hold by a small perturbation of our example if this perturbation is periodic with a sublattice of our original lattice of periods (of index two). 

\subsection{Counter-example}\label{4.1}

The following examle is due to N. Filonov (see \cite{FiKa}). We define the discrete Schr\"odinger operator in $l_2(\Z^2)$ as $H=\Delta+V$, where
$$
(\Delta u)_{(n_1,n_2)} = u_{(n_1+1,n_2)} +u_{(n_1-1,n_2)} +u_{(n_1,n_2+1)} +u_{(n_1,n_2-1)},
$$
and $(Vu)_{(n_1,n_2)}=V_0 u_{(n_1,n_2)}$ for $n_1+n_2$ being even and $(Vu)_{(n_1,n_2)}=V_1 u_{(n_1,n_2)}$ for $n_1+n_2$ being odd.

Then $H$ can be represented as the direct integral 
$$
H=\int\limits_{\tilde\Omega}^{\oplus}H({\bf k})\,d{\bf k},\ \ \ \ \ \tilde\Omega=\{{\bf k}\in\R^2:\ |k_1+k_2|\leq\pi\},
$$
where $H({\bf k})$ acts in $\C^2$ and is represented by the following matrix
\begin{equation*}
H({\bf k}):=\begin{pmatrix}
V_0 & 2\cos k_1+2\cos k_2 \\ 
2\cos k_1+2\cos k_2 & V_1
\end{pmatrix}.
\end{equation*}
Then it is easy to see that the spectrum of $H$ consists of two bands 
\bee
\left[\frac{V_0+V_1}{2}-\sqrt{\frac{(V_0-V_1)^2}{4}+16},\,\min\{V_0,V_1\}\right]
\ene 
and 
\bee
\left[\max\{V_0,V_1\},\,\frac{V_0+V_1}{2}+\sqrt{\frac{(V_0-V_1)^2}{4}+16}\right], 
\ene
and these intervals are disjoint unless $V_0=V_1$. Moreover, the upper edge of the first band and the lower edge of the second band both are attained on the boundary of $\tilde\Omega$. Thus, we have degenerate edges of the gaps with the corresponding degeneracy undestroyable with any small perturbation of the potential.

Of course, for the continuous Schr\"odinger operator the degeneracy on the lines is impossible by Thomas construction, moreover, as recently was proved in \cite{FiKa}, even degeneracy on the curves is impossible for $2$-dimensional continuous Schr\"odinger operators. However, this example is important as it shows that the question is not as obvious as it may look. Here we would also like to mention \cite{Sh} where the magnetic Schr\"odinger operator was constructed with degenerate lower edge of the spectrum (still attained at one point) while the proof from \cite{FiKa} excludes degeneracy on the curves for 2D magnetic operators too.

\ber
The example described above can be adjusted to obtain degeneracy even if the number of parameters is very large. For example, assume that $n\ge 3$ and consider the periodic operator with the lattice of periods generated by $(n,0)$  and $(1,1)$ and the potential $(V_0,V_1,\dots,V_{n-1})$ satisfying $V_0<V_j-2$. Then, if we put $s:=e^{ik_1}$ and $t:=e^{ik_2}$, 
the matrix of the fibre operator has the following form:
\begin{equation*}
H({\bf k}):=\begin{pmatrix}
V_0 & s+t & 0 & \dots & 0 & \overline {s+t} \\ 
\overline {s+t} & V_1 & s+t & 0 & \dots & 0 \\
\vdots & \vdots & \vdots & \vdots & \vdots&\vdots\\
s+t & 0 & \dots & 0 &\overline {s+t}&V_{n-1}
\end{pmatrix}.
\end{equation*}
Note that the quadratic form of $H({\bf k})$ equals $V_0$ on the vector $(1,0,\dots,0)$, which shows that the point $V_0$ is the right edge of the first spectral zone, attained at $\{s+t=0\}$. Thus the degeneracy of the spectral edge is an interval (not the union of two intervals as in the case $n=2$). 

\enr

\subsection{How to destroy the degeneracy by changing the lattice of periods}\label{4.2}
 
Now we discuss how the degeneracy in the example from the previous subsection will be destroyed using our approach. We consider the initial operator with doubled period (in vertical direction); we also assume (as we can without loss of generality) that  $V_0=V>0$, $V_1=-V$. Then we have 
$$
H=\int\limits_{\hat\Omega}^{\oplus}H({\bf k})\,d{\bf k},\ \ \ \ \ \hat\Omega:=\{{\bf k}\in\R^2:\ 0\leq k_j\leq\pi,\,j=1,2\},
$$
where $H({\bf k})$ acts in $\C^4$ and is represented by the following matrix
\begin{equation*}
H({\bf k}):=\begin{pmatrix}
V & 2\cos k_1 & 0 & 2\cos k_2 \\ 
2\cos k_1 & -V & 2\cos k_2 & 0 \\
0 & 2\cos k_2 & V & 2\cos k_1 \\
2\cos k_2 & 0 & 2\cos k_1 & -V
\end{pmatrix}.
\end{equation*}
We denote $a:=2\cos k_1$ and $b:=2\cos k_2$. The spectrum of the operator $H$ consists of two bands $[-\sqrt{V^2+16},-V]$ and $[V,\sqrt{V^2+16}]$. The edges $\pm V$ are attained when $a=\pm b$, i.e. on the diagonals of the square $\hat\Omega$. 

We will show that now the small perturbation of the potential destroys the degeneracy of the edges. We consider in details the lower edge of the second band, the construction for the upper edge of the first band is similar. Our perturbation has the form $B:=diag\{2\ep,0,0,0\}$, $\ep>0$. Equation $\det(H({\bf k})+B-\lambda)=0$ reads as follows
$$
[(\lambda^2-V^2)-(a^2+b^2)]^2-2\ep(\lambda+V)[(\lambda^2-V^2)-(a^2+b^2)]-4a^2b^2=0.
$$
We put $t:=[(\lambda^2-V^2)-(a^2+b^2)]$ and solve the quadratic equation for $t$. We get
\bee\label{tsquare}
t=\ep(\lambda+V)\pm\sqrt{\ep^2(\lambda+V)^2+4a^2b^2}.
\ene
First, let us show that the minimum of the second band is situated near the center of the square $a=b=0$. Indeed, obviously it must be near the diagonals $a=\pm b$. In a neighborhood of any point on the diagonals which is not the center of the square we have $|ab|\gg1$. Then \eqref{tsquare} gives
$$
(\lambda^2-V^2)-(a^2+b^2)=t=\ep(\lambda+V)\pm 2ab\left(1+O(\ep^2)\right)
$$
and thus
$$
\lambda^2-\ep\lambda -(a\pm b)^2-V^2-\ep V +O(\ep^2)=0.
$$
The eigenvalue corresponding to the second band is 
$$\lambda=\ep/2+\sqrt{V^2+\ep V+(a\pm b)^2+O(\ep^2)}\geq V+\ep+O(\ep^2).$$
At the same time, for $a=b=0$ there is the solution $t=0$ for \eqref{tsquare} which corresponds to $\lambda=V$, and thus, the lower edge of the second band occurs near the point $a=b=0$, i.e. $k_1=k_2=\pi/2$.

For the unperturbed operator we have $\lambda^2=V^2+(a\pm b)^2$. This leads to
$$
\la_\pm=V+\frac{1}{2V}\left(a\pm b\right)^2 +O((a^2+b^2)^2).
$$
We consider the two-dimensional subspace of the eigenvectors corresponding to the perturbed eigenvalues when $\ep\geq0$.
 Let $P$ be the orthogonal projection onto this subspace. We notice that while the eigenvectors, generally, are not analytic in $a,\ b$ and $\varepsilon$, the projection $P$ is analytic and can be represented by the convergent series $P=P_0+\sum_{n=1}^\infty \varepsilon^n P_n$ with $P_n=P_n(a,b)$ being analytic in the small neighborhood of $(0,0)$. Moreover, it is not hard to see that one can choose the analytic orthonormal basis  $f_1(a,b,\ep),\ f_2(a,b,\ep)$ in the range of $P$  such that $f_j(0,0,0)=(1,0,\pm 1,0)/\sqrt{2}$. Indeed, first we notice that the unperturbed matrix has analytic eigenvectors $f_j(a,b,0)$ with $f_j(0,0,0)=(1,0,\pm 1,0)/\sqrt{2}$ and then one should just apply Gram-Schmidt orthogonalization to $Pf_j(a,b,0)$. Since the range of $P$ is, obviously, an invariant subspace of $H(\bk)+B$, it is enough to consider the restriction of $H(\bk)+B$ to this space. The matrix of this restriction in the orthonormal basis constructed above has the following form: 
\begin{equation*}
M:=\begin{pmatrix}
\la_+ +\ep(1+g_1(a,b,\ep)) & \ep(1+g_3(a,b,\ep)) \\ 
\ep(1+g_3(a,b,\ep)) & \la_- +\ep(1+g_2(a,b,\ep))
\end{pmatrix},
\end{equation*}
with $g_j(0,0,0)=0$. 

Put $x=\frac{a+b}{2\sqrt{V}}$ and $y=\frac{a-b}{2\sqrt{V}}$. Then in these new coordinates the matrix $M-VI$ has the following  form: 
\bee
\tilde M:=\begin{pmatrix}
2\mu_1 & \sigma\varepsilon  \\ 
\sigma\varepsilon  & 2\mu_2
\end{pmatrix},
\ene
where $\mu_1=x^{2}+O((x^2+y^2)^2)+\varepsilon f_1(x,y,\varepsilon)$, $\mu_2=y^{2}+O((x^2+y^2)^2)+\varepsilon f_2(x,y,\varepsilon)$, $\sigma=1+f_3(x,y,\varepsilon)$. Functions $f_j$ are analytic in all variables and $f_3(0,0,0)=0$. Obviously, we are interested in the smallest eigenvalue of $\tilde M$, i.e.
$$
\tau:=\mu_1+\mu_2-\sqrt{(\mu_1-\mu_2)^2+\varepsilon^2 \sigma^2}.
$$
Let $(x_*,y_*)$ be a point of local minimum for $\tau$ in a small neighbourhood of zero, i.e. $|x_*|+|y_*|=o(1)$ as $\varepsilon\to0$. Without loss of generality we also assume $x_*^{2}\geq y_*^{2}$. We also notice that since the point $(0,0)$ was the minimum for the unperturbed eigenvalue, we also have $\mu_j(x_*,y_*)>-|o(\ep)|$.

Case 1. First, we assume that $\mu_1-\mu_2\geq 2\varepsilon$ at $(x_*,y_*)$. Then we notice that
\bee
\bes
&\tau(x_*,y_*)\geq (\mu_1-\mu_2)\left(1-\sqrt{1+\frac{\varepsilon^2 \sigma^2}{(\mu_1-\mu_2)^2}}\right)+o(\ep)\\
&\geq -\frac{\varepsilon^2 \sigma^2}{2(\mu_1-\mu_2)}+o(\ep)>-\varepsilon \sigma=\tau(0,0).
\end{split}
\ene
Thus, the lower edge of the zone is not attained at point $(x_*,y_*)$ and we can ignore this point.

Case 2. Let $\mu_1-\mu_2\leq 2\varepsilon$ at $(x_*,y_*)$. Then 
$$
1-\frac{\mu_1-\mu_2}{\sqrt{(\mu_1-\mu_2)^2+\varepsilon^2 \sigma^2}}\geq \frac{1}{20}.
$$
Now, direct calculation shows that
\bee
\bes
&0=\tau'_x(x_*,y_*)=\left(1-\frac{\mu_1-\mu_2}{\sqrt{(\mu_1-\mu_2)^2+\varepsilon^2 \sigma^2}}\right)(\mu_1)'_x\\
&+\left(1+\frac{\mu_1-\mu_2}{\sqrt{(\mu_1-\mu_2)^2+\varepsilon^2 \sigma^2}}\right)(\mu_2)'_x+\frac{O(\varepsilon^2)}{\sqrt{(\mu_1-\mu_2)^2+\varepsilon^2 \sigma^2}}\\
&=\left(1-\frac{\mu_1-\mu_2}{\sqrt{(\mu_1-\mu_2)^2+\varepsilon^2 \sigma^2}}\right)\left(2x_*(1+o(1))\right)
+O(\varepsilon)+\frac{O(\varepsilon^2)}{\sqrt{(\mu_1-\mu_2)^2+\varepsilon^2 \sigma^2}}\end{split}
\ene
and therefore 
\bee
x_*=O(\varepsilon)+\frac{O(\varepsilon^2)}{\sqrt{(\mu_1-\mu_2)^2+\varepsilon^2 \sigma^2}}=O(\varepsilon). 
\ene
 Thus, we proved that 
$$x_*=O(\varepsilon),\ \ \ \ \  y_*=O(\varepsilon).$$
Now direct calculation of the derivatives of $\tau$ gives 
\begin{equation}\label{secder1}
\frac{\partial^{2}\tau}{\partial x^{2}}(x_*,y_*)=2+O(\varepsilon),\ \ \ \frac{\partial^{2}\tau}{\partial y^{2}}(x_*,y_*)=2+O(\varepsilon),\ \ \ \frac{\partial^{2}\tau}{\partial x\partial y}(x_*,y_*)=O(\varepsilon).
\end{equation}

Thus, we have obtained the non-degenerate minimum.

\end{document}